\documentclass[aps, prd, twocolumn, amsmath, floats,floatfix, superscriptaddress, nofootinbib,longbibliography]{revtex4-2}

\usepackage{epsfig}
\usepackage{amsmath,amssymb,mathrsfs,mathtools}
\usepackage{verbatim}
\usepackage{physics}
\usepackage{float}
\usepackage{morefloats}
\usepackage[dvipsnames]{xcolor}
\usepackage{booktabs}
\usepackage[normalem]{ulem}
\usepackage{multirow}
\usepackage{makecell}
\usepackage{tabularx}
\usepackage{subcaption}
\usepackage{cancel}
\usepackage{orcidlink}
\usepackage[T1]{fontenc}
\usepackage[capitalise]{cleveref}
\usepackage{comment}

\crefname{section}{Sec.}{Secs.}
\crefname{table}{Tab.}{Tabs.}
\crefname{figure}{Fig.}{Figs.}
\crefname{equation}{Eq.}{Eqs.}
\crefname{appendix}{Appendix\ }{Appendix\ }

\providecommand{\openone}{\leavevmode\hbox{\small1\kern-3.8pt\normalsize1}}

\definecolor{bostonuniversityred}{rgb}{0.8, 0.0, 0.0}

\usepackage{xspace}
\usepackage{mathrsfs}

\DeclareSymbolFontAlphabet{\mathrsfs}{rsfs}

\usepackage{booktabs}
\usepackage[Q=yes,pverb-linebreak=no]{examplep}
\usepackage{aas_macros}
\usepackage{scrextend}

\allowdisplaybreaks[4]

\begin{document}

\title{\boldmath Rotating Thin Shells in Einstein-Gauss-Bonnet Gravity \unboldmath}

\author{Jo\~ao~D.~\'Alvares\,\orcidlink{0000-0001-5501-9014}}
\email{joaodinis01@gmail.com}
\affiliation{Department of Physics and Astronomy, University of Mississippi, University, MS 38677, USA}
\affiliation{Instituto de Telecomunicaç\~oes - IUL, Avenida das Forças Armadas, 1649-026 Lisboa, Portugal}

\author{Tiago V. Fernandes\,\orcidlink{0000-0001-6831-1137}}
\email{tiago.vasques.fernandes@tecnico.ulisboa.pt}
\affiliation{CENTRA, Departamento de F\'{\i}sica do Instituto Superior T\'{e}cnico (IST), Universidade de Lisboa, 1049-001 Lisboa, Portugal}
\affiliation{Departamento de Matemática, ISCTE - Instituto Universitário de Lisboa, Avenida das Forças Armadas, 1649-026, Lisboa, Portugal}


\begin{abstract}
A rotating metric solution in Einstein-Gauss-Bonnet gravity with a 
negative cosmological constant was 
recently found in the Chern-Simons point. We construct a rotating thin 
shell gluing two spacetimes in 
Einstein-Gauss-Bonnet gravity, using the Davis junction conditions. 
We take the inner and outer spacetimes as 
replicas of the same rotating metric, 
with different values of mass and angular momentum. We show that the only 
possible thin shells either are vacuum thin shells or have a non-zero 
pressure in one tangential direction while the remaining stress tensor 
components vanish. We obtain the equation of motion for the shell, which 
resembles the continuity equation for a shell in General Relativity (GR), 
even though the quantity analogous to the intrinsic mass of the shell in GR is 
not connected to its stress tensor. We study the special case of 
vacuum thin shells connecting two spacetimes with zero hair.
We obtain analytically the possible trajectories of the shell, and in certain 
situations we observe that the solution ceases to be valid. We find cases 
where the vacuum shell collapses and a naked singularity is formed.
There two types of static vacuum thin shell solutions, 
one being stable occurring when both inner 
and outer spacetimes are overextremal, and the other unstable occurring when 
the horizons of inner and outer spacetimes approach each other, and are close to 
extremality.

\end{abstract}

\maketitle

\section{Introduction}

Current observations have been consistent with the description of gravity 
given by General Relativity (GR). Since the inception of GR, modifications have been 
proposed to the theory~\cite{Weyl:1918ib}, for example, by including 
higher order curvature terms in the Lagrangian. While at the beginning the research 
around modified theories was driven by curiosity, it is now mainly motivated 
by curing the conceptual issues of GR. Namely, GR is incompatible with the 
Standard Model, and, furthermore, singularities occur in a wide range of GR 
solutions. These conceptual limitations may indicate that GR is the low-energy limit of a 
more general theory. It is recurrent to have contributions of higher order curvature terms 
in the Lagrangian from one-loop corrections of quantum fields~\cite{Birrell:1982ix} 
or even in the low energy regime of string theories~\cite{fradkin_effective_1985,boulware_string-generated_1985, Callan:1986jb,gross_quartic_1987}. A particular modified theory that arises from 
heterotic string theory is the dilatonic Einstein-Gauss-Bonnet gravity in five dimensions~\cite{gross_quartic_1987, Bento:1995qc}. As one works away from the cosmological 
scale, the dilaton field can be considered to be constant, yielding Einstein-Gauss-Bonnet 
(EGB) gravity. Since string theories are candidates for describing gravity at high energies, there has been considerable interest in studying EGB gravity. Moreover, EGB gravity in five dimensions corresponds to the five-dimensional Lovelock theory~\cite{lovelock_einstein_1971}, 
which has the property of having second-order field equations. Some connections between 
$d$-dimensional Lovelock theory and string theory have been made~\cite{Wang:2020eln}.
For a review of Lovelock gravity, see Ref.~\cite{garraffo_lovelock_2008}.

In general, Lovelock gravity has shortcomings regarding the number of 
degrees of freedom, as their number depends highly on the coefficients 
of the theory. There is a convenient choice of Lovelock coefficients that 
maximizes the number of degrees of freedom~\cite{TRONCOSO_2000,
Troncoso:1999pk}, which reduces the Lagrangian to the Chern-Simons 
form invariant under the anti-de Sitter (AdS) group in odd dimensions and 
to Born-Infeld densities in even dimensions. These particular cases 
of Lovelock gravity in odd dimensions can be used in the context of 
supergravity~\cite{TRONCOSO:1998ng}. Moreover, a great interest has been 
placed in the five-dimensional case
with negative cosmological constant, which is the Chern-Simons Einstein Gauss Bonnet 
AdS gravity or CS EGB AdS for short, 
due to the AdS/CFT conjecture~\cite{Maldacena:1997re}. As CS EGB AdS 
gravity constitutes the pure gravitational part of an AdS Chern-Simons 
supergravity~\cite{Zanelli:2005sa}, through the AdS/CFT correspondence, one is able 
to relate this theory to a conformal field theory at the boundary of AdS. One then can 
deduce properties of the states of the conformal field theory from the properties of 
solutions of EGB AdS gravity.

Regarding the Hilbert-Einstein AdS action, there are several black hole solutions 
with a spherical event horizon, namely the Kerr-Newman AdS family. 
Furthermore, the negative cosmological constant and higher dimensions allow the existence of 
event horizons with different topologies, namely, 
black branes, black strings, and other topological black holes. For EGB AdS gravity with general 
coupling, a spherically symmetric static solution similar to Schwarzschild AdS is the 
Boulware-Deser solution~\cite{boulware_string-generated_1985}, which is regular everywhere. This 
regularity may be interpreted as a short-distance effect of EGB, which can also be connected with 
low-energy effects of string theory as referred above. The Boulware-Deser 
solution has been extended to other topologies~\cite{cai_gauss-bonnet_2002}. Regarding 
rotating solutions, an analytic expression for charged rotating black branes has been 
found~\cite{dehghani_charged_2003,dehghani_magnetic_2004}, by using an ansatz of a boosted
planar black hole~\cite{Lemos:1995cm, Awad:2002cz}. Rotating solutions analogous to Kerr AdS 
have also been found, but only in the slow rotating approximation~\cite{PhysRevD.77.024045}
or numerically~\cite{Brihaye:2008kh}. In general, there is still no analogous solution to the Kerr-Newman solution in EGB gravity. An attempt was made 
using the Kerr-Schild ansatz, but the ansatz was shown not to solve 
the EGB equations for general coefficients~\cite{anabalon_kerr-schild_2009}, except for 
the special case of the Chern-Simons point.

In the particular Chern-Simons point, there are solutions that can be brought from the 
general EGB AdS gravity. For example, the Chern-Simons Deser-Boulware solution~\cite{Banados:1993ur}, 
its generalization to other topologies~\cite{Cai:1998vy,Aros:2000ij}, and the inclusion of 
electric charge~\cite{crisostomo_black_2000} have been found. Furthermore, 
solutions describing wormholes~\cite{Dotti:2006cp} and spacetime horns~\cite{Dotti:2007az} have been 
found. Considering the full CS supergravity instead of the gravitational part only, 
supersymmetric black hole analytic solutions also exist~\cite{Giribet:2014hpa}.
Regarding rotating solutions, these have been found numerically or through the slow rotating 
approximation in~\cite{brihaye_black_2013, kleihaus_phases_2023}. The analytical solutions 
with Kerr-Schild form found in~\cite{anabalon_kerr-schild_2009} for the Chern-Simons point 
are indeed rotating solutions but do not possess a Killing horizon and so do not seem to describe 
black holes. Fortunately, the endeavor to obtain analytical rotating black hole solutions was 
fruitful in~\cite{Anabalon:2024abz}, where the metric has a form related to the 
BTZ black hole~\cite{banados_black_1992}.

With the specific purpose of studying further the solution shown in~\cite{Anabalon:2024abz}, we turn now to a powerful tool to construct novel solutions by gluing patches of spacetimes. In GR, one can glue two spacetimes together as long as the Israel-Darmois 
junction conditions are obeyed~\cite{israel_singular_1966, darmois_equations_nodate}, see Ref.~\cite{Poisson_2004} 
for further details. 
These same conditions can be deduced through two methods: one 
using the properties of tensor distributions~\cite{israel_singular_1966} and 
the other using the variational principle in the Lagrangian formalism
with the Gibbons-Hawking-York boundary 
term of the action~\cite{Hayward:1990tz}. The variational principle for 
shells was extended also to the Hamiltonian formalism in Ref.~\cite{Hajicek:1997va}.
While these can be used to construct star models, they are mostly used 
to construct thin shells. Thin shells have been studied to understand gravitational 
collapse~\cite{Israel:1967zz, Kuchar:1968,delsate_collapsing_2014}, and to explore exotic solutions of GR. 
Namely, static shells around black holes have been considered~\cite{Brady:1991}, 
together with wormhole spacetimes~\cite{Bejarano:2011yz}, and tension shells~\cite{Katz:1991}.
Furthermore, regular black holes have also been constructed, see Ref.~\cite{Lemos:2021jtm} and references therein.

In EGB gravity, the junction conditions for gluing two spacetimes with a thin shell have been topic of debate. Using the variational principle with the 
boundary terms of the EGB action~\cite{Myers:1987yn}, Davis found 
the second junction conditions~\cite{Davis_2003,Ramirez:2024czb}, 
which we call the Davis junction 
conditions. However, using tensor distributions, 
the decomposition of the field equations contains quadratic Dirac delta 
terms which cannot be collected in a naive 
way~\cite{reina_junction_2016}. 
By making each quadratic term vanish, one has more restrictive junction 
conditions. Furthermore, there is the possibility of having dipole-type 
singularities in the field equations, giving rise to double layers. 
In~\cite{reina_junction_2016} and 
further correspondence with the authors, the collection of the quadratic Dirac delta 
terms may be implied~\cite{Davis_2003}, leading to their cancellation. In other words, the Davis junction conditions include a bigger space of solutions than the ones shown in Ref.~\cite{reina_junction_2016}, that agree with the Israel-Darmois conditions for GR. This might imply the existence of non-physical solutions~\cite{senovilla_junction_2026}. In~\cite{Chu:2021uec}, 
a treatment regarding Dirac delta as a limit of regular distributions was considered to avoid 
the ill-defined sums of quadratic Dirac delta terms and other singularities, and 
the junction conditions obtained coincided with~\cite{Davis_2003}. To our knowledge, 
this debate is still not settled, and further comparison between tensor distributions and the variational 
principle is needed.

The Davis junction conditions are also obeyed for $C_2$ metrics, 
which are used to construct star models in 
EGB gravity~\cite{Ilha:1999yn,brassel_stars_2023}. 
Moreover, thin shell solutions have also been studied, e.g., 
charged thin shells through the Hamiltonian formalism~\cite{dias_charged_2007}, collapsing thin shells~\cite{crisostomo_hamiltonian_2004, Huang:2021bdm}, wormhole thin shells~\cite{Eiroa:2025ddo}, 
membranes with intrinsic charges~\cite{Liu:2024aos}, and thin shells 
describing dark halos~\cite{chahboun_einstein-gauss-bonnet_2024}, all of 
them relying on the Davis junction conditions.
An interesting type of thin shell solution has been found in EGB gravity, describing 
a dynamic surface with vanishing surface stress tensor. These solutions are named 
vacuum thin shells and have been studied in~\cite{gravanis_mass_2007,garraffo_gravitational_2008,ramirez_vacuum_2018,
Ramirez_2024, garrafo_2010}.

In this paper, we study thin shell solutions that glue two patches of spacetimes 
described by the recently found rotating solutions in CS EGB AdS gravity~\cite{Anabalon:2024abz}. 
By using the Davis junction conditions, we obtain the 
form of the surface stress tensor that can solve the conditions. Moreover, we study 
the case of vacuum thin shells around a black hole with no hair and obtain analytical 
expressions for the motion of the shells. We also find stationary vacuum thin shell solutions 
and analyze their stability. \par
This paper is structured as follows: we start by discussing CS EGB AdS gravity 
and its recent solution~\cite{Anabalon:2024abz} in Sec.~\ref{sec:EGB}.
We then present the junction conditions in Sec.~\ref{sec:junc_cond}, 
followed up by the equations of motion that these same conditions provide in Sec.~\ref{sec:eomshell}. 
We then proceed to analyze vacuum thin shells in Sec.~\ref{sec:vacuumthinshells}. 
Some final remarks are discussed in Sec.~\ref{sec:conclusions}.
We will use natural units throughout the entirety of this paper ($G = c = 1$).

\section{Rotating black hole solution in Chern-Simons Einstein-Gauss-Bonnet gravity}
\label{sec:EGB}
\subsection{Chern Simons Einstein-Gauss-Bonnet gravity}

Lovelock-Lanczos gravity~\cite{lovelock_einstein_1971} 
is the most general theory built from the metric only where the field equations 
are of second order. Its bulk action is $S_{\mathrm{lov}} = \int \mathcal{L}_\mathrm{lov} d^d x$, 
with the Lagrangian
\begin{equation}
    \mathcal{L}_\mathrm{lov} = \sqrt{-g}\sum_{n=0}^t \alpha_n \mathrm{R}^n\,,
    \label{eq:lagGeneral}
\end{equation}
where $t$ represents the number of terms being taken into account, $\alpha_n$ are scalar constants, the $\mathrm{R}^n$ are defined as
\begin{equation}
    \mathrm{R}^n = \frac{1}{2^n}\delta^{\mu_1\nu_1 ... \mu_n \nu_n}_{\alpha_1\beta_1 ... \alpha_n \beta_n}\prod_{r=1}^n {R_{\mu_r \nu_r}}^{\alpha_r \beta_r}\,.
\end{equation}
Note that $\delta^{\mu_1\nu_1 ... \mu_n \nu_n}_{\alpha_1\beta_1 ... \alpha_n \beta_n}$ vanishes  trivially for $n>d$, where $d$ is the number of dimensions of the 
spacetime~\cite{lovelock_einstein_1971}, ${R_{\mu_r \nu_r}}^{\alpha_r \beta_r}$ is the Riemann tensor given by the metric $g_{\mu \nu}$ and 
its first and second derivatives, and $g$ is the determinant of the metric. 
The first term of the sum in Eq.~\eqref{eq:lagGeneral} corresponds to $\alpha_0$, i.e., a cosmological constant, 
while the second term is the Einstein-Hilbert term, 
and the third one is known as the Gauss-Bonnet term. Knowing that the number of terms is determined by the spacetime dimensions, 
we get that in $d=4$ only the first two terms are possible. Hence, the third term only enters for 
$d\geq 5$. Curiously, the Gauss-Bonnet term corresponds to the Euler density in four dimensions, and so EGB in four dimensions 
reduces to GR. Explicitly, the five-dimensional EGB AdS gravity has the action 
\begin{equation}\label{eq:action}
    S = S_{\text{EH}}+S_{\text{GB}}
\end{equation}
where 
\begin{align}
    S_{\text{EH}} &= \frac{1}{16\pi}\int d^5 x \sqrt{-g}(R-2\Lambda) - \frac{1}{8\pi}\int d^4 x \sqrt{-h}K\,\,\\
    S_{\text{GB}} &= \frac{\alpha}{16\pi}\int d^5x \sqrt{-g}\left(R^2 - 4R_{ab}R^{ab} + R^{abcd}R_{abcd} \right)\notag\\
        &-\frac{\alpha}{4\pi}\int d^4 x \sqrt{-h}\left(J - 2 \widehat{G}^{ij}K_{ij} \right)\,,
\end{align}
with $\alpha$ being the Gauss-Bonnet coefficient, $\Lambda$ being the cosmological constant, $h_{ij}$ the induced metric on the boundary, 
$h$ the determinant of the induced metric; $K_{ij} = e_i^a e_j^b \nabla_a n_b$ is the extrinsic curvature at the boundary, 
being $n_a$ the normal covector to the boundary, $\widehat{G}_{ij}$ is the Einstein tensor constructed from the induced metric, 
and $J = J_{ij} h^{ij}$ is the trace of $J_{ij}$, which is described by
\begin{equation}
    \begin{aligned}
        J_{ij} = \frac{1}{3}\Big(2K K_{ik}K^k_j &+ K_{kl}K^{kl}K_{ij} - \\
        &-2K_{ik}K^{kl}K_{lj} - K^2 K_{ij}\Big)\,.
    \end{aligned}
\end{equation}
By comparison with Eq.~\eqref{eq:lagGeneral}, EGB AdS is recovered by setting $\alpha_0 = -2\Lambda$, $\alpha_1 = 1$, and $\alpha_2 = \alpha$.
Note that we have introduced the boundary terms of the action as derived in Ref.~\cite{Myers:1987yn}, which are important 
when we consider thin shells. Moreover, GR with a negative cosmological constant can be recovered by setting
$\alpha=0$. 

Here, we are interested in CS EGB AdS gravity, which is a particular case of 
EGB AdS with the coefficient $\alpha$ related to the cosmological constant 
by $\alpha \Lambda = - \frac{3}{4}$. 
We define the AdS length in five dimensions as 
\begin{align}
    l^2 = - 3\Lambda\,\,,
\end{align}
and so the Gauss-Bonnet coefficient is given by
\begin{align}
    \alpha = \frac{l^2}{4}\,\,.
    \label{eq:chernsimons}
\end{align}
This relation between $\alpha$ and $\Lambda$ holds specifically only for $d=5$ dimensions~\cite{anabalon_kerr-schild_2009}.

\subsection{Rotating solution}

The black hole rotating solution in CS EGB AdS gravity 
found in~\cite{Anabalon:2024abz} has a metric   
\begin{equation}
    \begin{aligned}
        ds^2 &= l^2 \cosh^2(\rho)\left[-A(r)dt^2 + \frac{dr^2}{A(r)} + r^2 (d\psi + N^\psi dt)^2 \right]\\
        &+l^2 d\rho^2 + l^2 \cosh^2(\rho-\rho_0)dz^2\,,
    \end{aligned}
    \label{eq:Anabalon}
\end{equation}
where
\begin{equation}
    A(r) = r^2 - M - \frac{b}{r} + \frac{j^2}{4r^2}\,, \quad N^\psi = -\frac{j}{2r^2}\,,
    \label{eq:NandNpsi}
\end{equation}
written in Boyer-Lindquist type coordinates $(t,r, \psi, \rho, z)$, with 
$\psi \in [0,2\pi[$ being an angular coordinate. The metric is parametrized 
by three parameters, $M$ being the mass, $j$ being the angular momentum and 
$b$ being a hair parameter. The spacetime has an event horizon located at 
$r=r_+$, and a Cauchy horizon at $r=r_-$, that can be deduced from Eq.~\eqref{eq:NandNpsi}. When $b=0$, the expressions for both horizons simplify to
\begin{align}
    r_\pm^2 = \frac{M\pm \sqrt{M^2 - j^2}}{2}\,\,.
\end{align}
It should also be noted that, when $b = 0$, the metric given by Eq.~\eqref{eq:Anabalon} stops having intrinsic rotation.

\section{Thin shells in Einstein-Gauss-Bonnet and their junction conditions}
\label{sec:junc_cond}
We are interested in constructing a thin shell, $\Sigma$, 
that glues two spacetimes described by Eq.~\eqref{eq:Anabalon}, 
say $\mathcal{M}_-$ and $\mathcal{M}_+$,
with different sets of parameters $M$, $j$, $\rho_0$ and $b$. The quantities inside the shell will 
have a minus sign ($M_-$, $j_-$, $\rho_{0_-}$, $b_-$) for $\mathcal{M}_-$, while 
the ones outside the shell will be 
accompanied by a plus sign ($M_+$, $j_+$, $\rho_{0_+}$, $b_+$) for $\mathcal{M}_+$. 
The same follows for all geometric quantities.\par
The metric then needs to satisfy the junction conditions where both spacetimes are glued. 
We will use the Davis junction conditions~\cite{Davis_2003}. 
The first condition is the same as for the Darmois-Israel formalism, i.e., that the induced metrics $h_{ij}^\pm$ 
computed from $g_{\mu\nu}^+$ and from $g_{\mu\nu}^-$ must be continuous across the thin shell $\Sigma$,
\begin{equation}
    h_{ij}^+ = h_{ij}^- = h_{ij}\,,
    \label{eq:1JunctionCondition}
\end{equation}
where we introduce $h_{ij}$ to simplify notation. The other junction condition 
comes from the variation of the action, Eq.~\eqref{eq:action}, in the induced metric 
of the thin shell. This condition
relates the jump of the extrinsic curvature across $\Sigma$ with the 
surface stress tensor of the thin shell as
\begin{equation}
    [K_{ij}-K h_{ij}]+2\alpha [3J_{ij} - Jh_{ij} + 2\widehat{P}_{iklj}K^{kl}] = -8\pi S_{ij}
    \label{eq:2JunctionCondition}
\end{equation}
where $P_{abcd}$ is the divergence-free Riemann tensor,
\begin{equation}
    P_{abcd} = R_{abcd} + 2 R_{b[c}g_{d]a}-2R_{a[c}g_{d]b}+Rg_{a[c}g_{d]b}\,,
\end{equation}
and $\widehat{P}_{ijkl}$ meaning that it is constructed from the induced quantities. 
$S_{ij}$ is the stress-energy tensor of the thin shell, and the square brackets enclosing 
a given quantity $A$ correspond to the jump of $A$ across the thin shell, 
i.e., $[A] = A_+ - A_-$. This definition is different from Ref.~\cite{Davis_2003}, 
only due to the fact that we calculate the extrinsic curvature from either side of 
the shell with the same normal vector, as is more usually done.

\subsection{First Junction Conditions}
\label{sec:firstjunccond}
Before addressing the first junction conditions, we perform two changes of variables. 
The first of these changes will be done in order to remove the off-diagonal 
term of the metric by changing to the comoving frame, as done in Refs.~\cite{PhysRevD.79.064005, delsate_collapsing_2014},
\begin{equation}
    d\psi \rightarrow d\phi - \Omega(t) dt\,,
    \label{eq:anglechange}
\end{equation}
where $\Omega(t)$ is a function of time that we will choose appropriately. Let also the thin shell $\Sigma$ be defined as $\Sigma = \{x^\mu : t = \mathcal{T}(\tau), r = \mathcal{R}(\tau) \}$. This means that the coordinate system of the shell is given by $y^i = \{\tau, \phi, \rho, z\}$. It has become common practice to define $\tau$ to be the proper time, but this choice will make the rest of the derivation inconsistent. Instead, we choose $\tau$ such that Eq.~\eqref{eq:Anabalon}, together with Eq.~\eqref{eq:anglechange}, becomes
\begin{equation}
    \begin{aligned}
        ds^2 = l^2 \cosh^2(\rho)&\left[-d\tau^2 + \mathcal{R}^2(\tau)d\psi^2\right]+\\
        &+l^2 d\rho^2 + l^2 \cosh^2(\rho-\rho_0)dz^2\,.
    \end{aligned}
    \label{eq:AnabalonProperTime}
\end{equation}
The new time coordinate $\tau$ can be thought of as the proper time of the BTZ part of the metric. Eq.~\eqref{eq:AnabalonProperTime} implies that
\begin{subequations}
    \begin{align}
    \dot{\mathcal{T}}(\tau) = &\frac{\sqrt{A(\mathcal{R}(\tau)) + \dot{\mathcal{R}}^2(\tau)}}{A(\mathcal{R}(\tau))}\\
    \Omega(\tau) = N^\psi (\tau)& \frac{\sqrt{A(\mathcal{R}(\tau)) + \dot{\mathcal{R}}^2(\tau)}}{A(\mathcal{R}(\tau))} \,,
    \end{align}
    \label{eq:TOmegacond}
\end{subequations}
where the dots appearing on the left-hand side of the equations mean the derivative with respect to $\tau$. The procedure followed up until this point can be applied to the inner and the outer spacetimes, hence we have refrained from adding any plus or minus signs. The first junction condition, Eq.~\eqref{eq:1JunctionCondition}, is automatically fulfilled if:
\begin{equation}
    \begin{aligned}
        \{\tau_+, \phi_+,\rho_+, z_+\} &= \{\tau_-, \phi_-,\rho_-, z_-\}\\
        \mathcal{R}_+(\tau) &= \mathcal{R}_-(\tau)\\
        \rho_{0_+} &= \rho_{0_-}\,.
    \end{aligned}
\end{equation}
Hence, we will drop the plus and minus signs on these quantities from now on. It is remarkable that there is no restriction on the angular momenta, as was seen in Ref.~\cite{delsate_collapsing_2014}, for example.

\subsection{Second Junction Conditions}
\label{sec:secondjunccond}
Regarding the second junction conditions, Eq.~\eqref{eq:2JunctionCondition}, the non-vanishing entries of the extrinsic curvature are the following (apart from symmetries):
\begin{equation}
    \begin{aligned}
        K_{\tau\tau} &= -\frac{l \cosh(\rho)}{2}\frac{A^\prime(\mathcal{R}) + 2 \Ddot{\mathcal{R}}}{\sqrt{A(\mathcal{R})+\dot{\mathcal{R}}^2}}\\
        K_{\tau \phi} &= \frac{l \cosh(\rho)}{2}\mathcal{R}^2 (N^\psi(\mathcal{R}))^\prime\\
        K_{\phi\phi} &= l\cosh(\rho) \mathcal{R}\sqrt{A(\mathcal{R})+\dot{\mathcal{R}}^2}\,,
    \end{aligned}
\end{equation}
where $^\prime$ means the derivative with respect to $\mathcal{R}$. $J_{ij}$ is even simpler, with all its entries vanishing. Lastly, the divergence-free Riemann tensor constructed from the induced metric $\widehat{P}_{ijkl}$ is described by the following entries:
\begin{align}
        \hat{P}_{\rho z z \rho} &= \frac{l^2 \cosh(\rho-\rho_0)^2(\mathcal{R}\tanh(\rho)^2 - \text{sech}(\rho)^2 \Ddot{\mathcal{R}})}{\mathcal{R}}\nonumber\\
        \hat{P}_{\phi z z \phi} &= l^2 \cosh^2(\rho) \cosh(\rho-\rho_0)^2 \mathcal{R}^2\nonumber\\
        \hat{P}_{\tau\phi\tau\phi} &= l^2 \cosh^2(\rho) \mathcal{R}^2\\
        \hat{P}_{\tau\rho\tau\rho} &= l^2 \cosh(\rho) \sinh(\rho) \tanh(\rho-\rho_0)\nonumber\\
        \hat{P}_{\tau z \tau z} &= l^2 \cosh^2(\rho) \cosh(\rho-\rho_0)^2\nonumber\,.
\end{align}
The left-hand side (LHS) of the second junction conditions, Eq.~\eqref{eq:2JunctionCondition}, that we will call $E_{ij}$, has only one non-zero component:
\begin{equation}
    E_{\rho\rho} = \frac{l \cosh(\rho_0)}{\cosh^2(\rho) \cosh(\rho-\rho_0) \mathcal{R} \dot{\mathcal{R}}}\dot{m}\,,
    \label{eq:lhs}
\end{equation}
where we have defined $m(\tau)$ as
\begin{equation}
    m(\tau) = \mathcal{R} \left( \sqrt{A_-(\mathcal{R}) + \dot{\mathcal{R}}^2}-\sqrt{A_+(\mathcal{R}) + \dot{\mathcal{R}}^2}\right)\,.
    \label{eq:massdef}
\end{equation}
We give the name $m$ to Eq.~\eqref{eq:massdef} because it is a quantity that will appear together with the mass parameters, giving an idea of the discontinuity between the inner and outer spacetimes. This quantity would correspond to the $\beta_\pm$ in Ref.~\cite{delsate_collapsing_2014}.

\section{Equations of Motion of the Thin Shell}
\label{sec:eomshell}
\subsection{Thin Shell with an Anisotropic Massless Fluid}
\label{sec:anisotropicfluid}

A possible source of Eq.~\eqref{eq:lhs} is anisotropic fluids~\cite{delsate_collapsing_2014}, where the useful anisotropy for our case will be a different pressure along the $\rho$ direction. The stress-energy tensor of the thin shell in this case can be written as
\begin{equation}
    S_{ij} = (\epsilon + P_1) u_i u_j + P_1 (h_{ij} + w_i w_j) + P_2 w_i w_j\,,
\end{equation}
where $\epsilon$ is the energy density, $P_1$ is the pressure perpendicular to $\rho$, $P_2$ is the pressure along $\rho$, $u_i = (l \cosh (\rho),0,0,0)$ is the fluid four-velocity and $w_i = (0,0,l,0)$. We now must replace this stress-energy tensor into Eq.~\eqref{eq:2JunctionCondition}, and also check the conservation equation $D^i S_{ij} = 0$.
Because the only non-zero entry of the LHS of Eq.~\eqref{eq:2JunctionCondition} is $E_{\rho\rho}$, it follows immediately that $\epsilon = P_1 = 0$. Equation~\eqref{eq:lhs} can then only be satisfied if
\begin{equation}
    \frac{l \cosh(\rho_0)}{\cosh^2(\rho) \cosh(\rho-\rho_0) \mathcal{R} \dot{\mathcal{R}}}\dot{m} = -8\pi l^2 P_2\,,
    \label{eq:P2cond}
\end{equation}
while the conservation equation $D^i S_{ij} = 0$ yields
\begin{equation}\label{eq:conservationshell}
    \partial_\rho P_2 + \left(2 \tanh (\rho) + \tanh(\rho-\rho_0) \right) P_2 = 0\,.
\end{equation}
Notice that $P_2$ has to be a function of $\rho$ and $\tau$ for consistency. The explicit dependence on $\rho$ can be found by integrating Eq.~\eqref{eq:conservationshell}, giving 
\begin{equation}
    P_2(\rho, \tau) = \frac{\cosh(\rho_0)}{l\cosh^2(\rho) \cosh(\rho-\rho_0)}p(\tau)\,,
    \label{eq:pressurerho}
\end{equation}
where we have defined $p(\tau)$ as being the time dependence of the intrinsic pressure. 
Substituting this expression in Eq.~\eqref{eq:P2cond}, the junction condition reduces to
a differential equation depending only on the coordinate $\tau$ as
\begin{equation}
    \dot{m} = -8\pi \mathcal{R} \dot{\mathcal{R}} p\,.
    \label{eq:secondEquatioOfMotion}
\end{equation}
The equation of motion of the shell can then be obtained from Eq.~\eqref{eq:secondEquatioOfMotion} together with Eq.~\eqref{eq:massdef}.

The expression in Eq.~\eqref{eq:secondEquatioOfMotion} coincides with the conservation equation in the case of a shell in Schwarzschild spacetime in GR~\cite{Brady:1991}, where $m = 4\pi \mathcal{R}^2 \epsilon$ has the meaning of the intrinsic mass of the shell. Here, however, the quantity $m$ in Eq.~\eqref{eq:massdef} is not the intrinsic mass of the shell, since $\epsilon = 0$. 
Indeed, $m$ is just an auxiliary quantity that we defined. How can $m$ be interpreted here then? Due to the Gauss-Bonnet term, the junction condition in Eq.~\eqref{eq:2JunctionCondition} suffers a modification that cancels the term from GR in the $\tau \tau$ component. More precisely, it is the $\tau\tau$ component in GR that matches the right-hand side of Eq.~\eqref{eq:massdef} to the intrinsic energy of the shell. One can then interpret that the origin of the ``energy'', $m$, that allows for the gluing of these spacetimes, lies in the Gauss-Bonnet action, i.e., $m$ has its origin in the curvature rather than the shell.

\subsection{Equation of Motion}
\label{sec:equationsofmotion}

We can write the equations of motion to have an explicit relation between the derivatives of $\mathcal{R}$ and the pressure $p(\tau)$. By taking the derivative with respect to $\tau$ of Eq.~\eqref{eq:massdef}, we get
\begin{equation}\label{eq:mdot}
    \dot{m} = - 8 \pi \mathcal{R} \dot{\mathcal{R}} p_\mathcal{R} - \frac{m \dot{\mathcal{R}} \ddot{\mathcal{R}}}{\sqrt{(A_++\dot{\mathcal{R}}^2)(A_-+\dot{\mathcal{R}}^2)}}\,,
\end{equation}
where $p_\mathcal{R}$ is a functional defined by
\begin{equation}
    \begin{aligned}
        p_\mathcal{R} = \frac{1}{16 \pi \mathcal{R}} \hskip-0.5mm\Bigg(\hskip-0.5mm\frac{\mathcal{R} A_+^\prime + 2 A_+ + 2 \dot{\mathcal{R}}^2}{\sqrt{A_+ +\dot{\mathcal{R}}^2}} -\frac{\mathcal{R} A_-^\prime + 2 A_- + 2 \dot{\mathcal{R}}^2}{\sqrt{A_- +\dot{\mathcal{R}}^2}}\hskip-1mm\Bigg)\,,
    \end{aligned}
    \label{eq:PrDef}
\end{equation}
which looks like a pressure term that comes purely from curvature. With this definition, Eq.~\eqref{eq:secondEquatioOfMotion} becomes
\begin{equation}
    \Ddot{\mathcal{R}} = -\frac{8\pi \mathcal{R}}{m}(p_\mathcal{R}-p)\sqrt{(A_- + \dot{\mathcal{R}}^2)(A_+ + \dot{\mathcal{R}}^2)}\,\,,
    \label{eq:rdotdot}
\end{equation}
where $m$ is given by Eq.~\eqref{eq:massdef}. The equation of motion of the shell, Eq.~\eqref{eq:rdotdot}, is a nonlinear second-order ordinary differential equation which can be integrated by providing the initial radius and radial velocity of the shell. Note, however, that an analysis of the uniqueness of the solution is involved without the specification of the intrinsic pressure of the shell, $p$. Nevertheless, the 
differential equation only makes sense when 
$(A_+ + \dot{\mathcal{R}}^2)> 0$ and 
$(A_- + \dot{\mathcal{R}}^2) > 0$, otherwise $\dot{m}$ in 
Eq.~\eqref{eq:mdot} is ill-defined.

\subsection{Static Solutions and Stability Analysis: General Considerations}
\label{sec:staticsols}

It is convenient to explore the existence of static solutions of Eq.~\eqref{eq:rdotdot}. If such solutions exist, we then have
$\Ddot{\mathcal{R}} = \dot{\mathcal{R}} = 0$. Therefore, Eq.~\eqref{eq:rdotdot} is solved only if
\begin{equation}\label{eq:balancepressures}
    p_\mathcal{R} = p\,,
\end{equation}
for a configuration with nonzero $m$. This condition means that the shell can be static if and only if there is a balancing of the pressures at play in this scenario. 

Assuming that Eq.~\eqref{eq:balancepressures} is satisfied for a given radius, we can treat the stability of the static shells. To do this, we do a perturbation in $\mathcal{R} \rightarrow \mathcal{R}_0 + \delta \mathcal{R}$, and keep everything at lowest order. All quantities with a subscript 0 refer to the static shell. Applying this procedure to Eq.\eqref{eq:rdotdot}, we get
\begin{equation}
    \Ddot{\delta\mathcal{R}} = -\frac{8\pi \mathcal{R}_0}{m_0}\sqrt{A_- A_+}\partial_\mathcal{R}(p_\mathcal{R}-p)\big|_0\,,
\end{equation}
noting that $\partial_{\dot{\mathcal{R}}} p_\mathcal{R} \big|_0 = 2 \mathcal{R}_0 \partial_{\dot{\mathcal{R}}^2} p_\mathcal{R} \big|_0 = 0$, with $\dot{\mathcal{R}}_0 = 0$. Therefore, the static shell is stable if and only if
\begin{equation}\label{eq:stability}
    \partial_\mathcal{R}(p_\mathcal{R}-p)\big|_0 > 0\,.
\end{equation}

For any given $p$, one can now apply Eq.~\eqref{eq:balancepressures} to find the static solutions and use Eq.~\eqref{eq:stability} to probe for the stability of such solutions.

\section{Vacuum Thin Shells}
\label{sec:vacuumthinshells}

\subsection{Preamble}

We now analyze thin shells that have vanishing 
stress energy tensor, which occur in EGB~\cite{gravanis_mass_2007,garraffo_gravitational_2008,
ramirez_vacuum_2018,
Ramirez_2024}, but have no GR analogues. 

Vacuum thin shells have $S_{ij} = 0$, hence the quantity $p$ vanishes and so the equation of motion of the shell, Eq.~\eqref{eq:secondEquatioOfMotion}, becomes $\dot{m} = 0$. This can be integrated into $m(\tau) = m$, i.e., $m$ is a constant of motion of the shell. Using Eq.~\eqref{eq:massdef}, this condition can be rewritten to
obtain $\dot{\mathcal{R}}$ as 
\begin{equation}
    \dot{\mathcal{R}}^2(\tau) + V(\tau, m) = 0\,,
    \label{eq:motionpotential}
\end{equation}
where 
\begin{equation}
    \begin{aligned}
        &V(\tau) = -\left(m^2 - \mathcal{R}^2(\tau)\left(\sqrt{A_+} - \sqrt{A_-}\right)^2 \right) \cdot\\
        &\cdot \left(m^2 - \mathcal{R}(\tau)^2 \left(\sqrt{A_+} + \sqrt{A_-}\right)^2 \right)\frac{1}{4 \mathcal{R}^2(\tau)m^2}\,.
    \end{aligned}
\end{equation}
It is important to note that the second-order differential equation $\ddot{\mathcal{R}} = \ddot{\mathcal{R}}(\mathcal{R},\dot{\mathcal{R}})$, solved by giving the initial radius and radial velocity, becomes now a branched first order differential equation described 
by Eq.~\eqref{eq:motionpotential}, i.e., $\dot{\mathcal{R}} = \sqrt{- V(\mathcal{R}(\tau),m)}$ and $\dot{\mathcal{R}} = -\sqrt{- V(\mathcal{R}(\tau,m))}$. These equations are solved for an initial $\mathcal{R}_0$, while
the constant $m$ establishes the initial value of $|\dot{\mathcal{R}}_0|$ and the branch is chosen according to the sign of the initial value of $\dot{\mathcal{R}}_0$. This may complicate the 
analysis, since the solutions change branch when they reach $\dot{\mathcal{R}}_0=0$. Nevertheless,
this analysis can be helpful to find analytic solutions when the second-order equation is quite involved.

An important comment has to be made regarding the order reduction above. In order to invert Eq.~\eqref{eq:massdef}, we have to perform the squaring of both 
sides twice. In practice, this means that Eq.~\eqref{eq:motionpotential} 
can describe another condition 
$m = m_2(\tau)$, with $m_2$ defined below. In fact, this can be verified by solving Eq.~\eqref{eq:motionpotential} for $m$, thus obtaining Eq.~\eqref{eq:massdef}, and its symmetric, as well as 
$m = m_2(\tau)$ with 
\begin{align}\label{eq:otherm}
    m_2(\tau) = \mathcal{R}(\tau)\left(\sqrt{A_- + \dot{\mathcal{R}}^2} 
    + \sqrt{A_+ + \dot{\mathcal{R}}^2}\right)\,\,,
\end{align}
and its symmetric. When considering the initial conditions, $m$ must be calculated with Eq.~\eqref{eq:massdef}, which avoids this problem. However, it may happen that when a point with $A_\pm + \dot{\mathcal{R}}^2= 0$ is reached, then the analytical solution obeying Eq.~\eqref{eq:motionpotential} begins to describe 
Eq.~\eqref{eq:otherm} rather than Eq.~\eqref{eq:massdef}, continuing the solution. 
In comparison, evolving the differential equation in Eq.~\eqref{eq:secondEquatioOfMotion} 
beyond such a point may lead to $A_\pm + \dot{\mathcal{R}}^2< 0$ thus stopping the evolution.

We therefore restrict the parameter $m$ as defined originally by Eq.~\eqref{eq:massdef} to be real in general (i.e., avoiding any square roots to take on imaginary values). This places some restrictions on the possible values of $\mathcal{R}$ and $\dot{\mathcal{R}}$ for the shell, as the values inside the square roots of Eq.~\eqref{eq:massdef} have to be non-negative. Note that one can still have a shell shrink beyond the event horizon as long as the value of $\dot{\mathcal{R}}$ is large enough. Moreover, it is possible to consider an outside spacetime described by a super-extremal metric. 

We now split the analysis into the $b_\pm = 0$ and $b_\pm \neq 0$ cases. 

\subsection{Analytic solutions for $b_\pm = 0$}

\subsubsection{Dynamic solutions}

When both $b_\pm = 0$, there are several analytic solutions that Eq.~\eqref{eq:motionpotential} can take, depending on the initial data. Writing Eq.~\eqref{eq:motionpotential} explicitly, as a function of the following quantities,
\begin{equation}\label{eq:masterquantities}
    \begin{aligned}
        Q_0 = & (j_-^2 - j_+^2)^2 + 8m^2 (2m^2 - (j_+^2 + j_-^2))\\
        Q_2 = &8\left((j_+^2-j_-^2)(M_--M_+)+ 4m^2(M_-+M_+)\right)\\
        Q_4 = & 16(4m^2-(M_--M_+)^2)\,,
    \end{aligned}
\end{equation}
we get
\begin{equation}
    \dot{\mathcal{R}} = \pm\frac{\sqrt{Q_0 + Q_2 \mathcal{R}^2 - Q_4 \mathcal{R}^4}}{8m\mathcal{R}}\,,
    \label{eq:rdotanalytical}
\end{equation}
where the subscript of each $Q_i$ is related to the power of $\mathcal{R}$ that it is multiplying by.

\begin{figure} 
\centering 
\begin{subfigure}[t]{0.48\columnwidth} 
\centering \includegraphics[width=\textwidth]{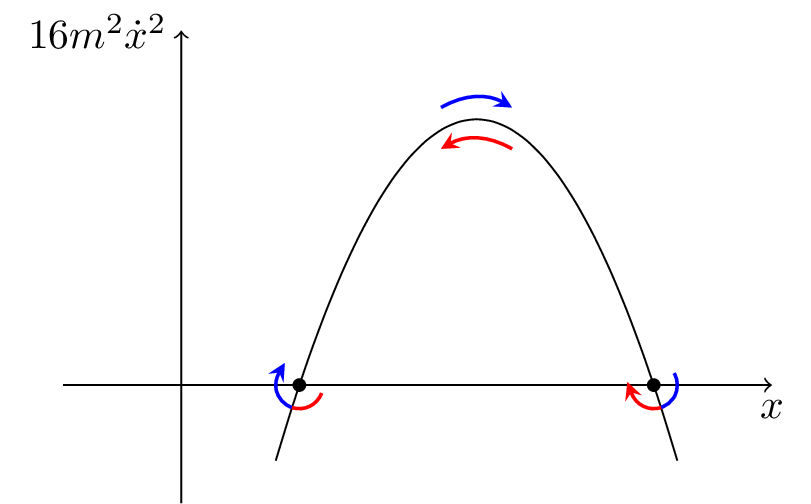} \caption{\label{fig:phasespaceQ4pdeltap}$Q_4 > 0$ and $\Delta > 0$} \end{subfigure} \begin{subfigure}[t]{0.48\columnwidth} \centering \includegraphics[width=0.8\textwidth]{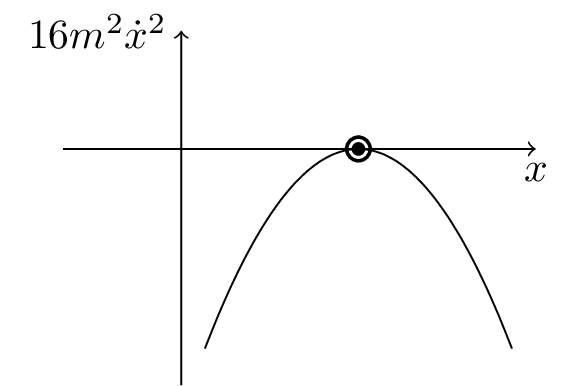} \caption{\label{fig:phasespaceQ4pdelta0}$Q_4 > 0$ and $\Delta = 0$} \end{subfigure} \begin{subfigure}[t]{0.48\columnwidth} \centering \includegraphics[width=\textwidth]{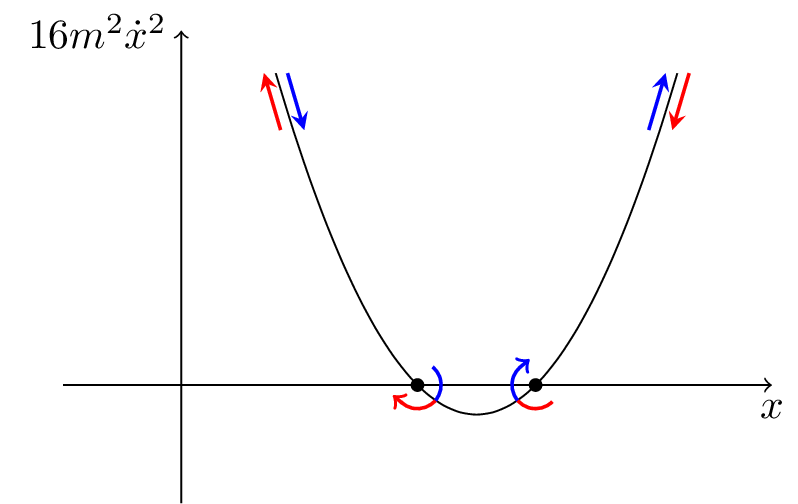} \caption{\label{fig:phasespaceQ4ndeltap}$Q_4<0$ and $\Delta > 0$} \end{subfigure} \begin{subfigure}[t]{0.48\columnwidth} \centering \includegraphics[width=\textwidth]{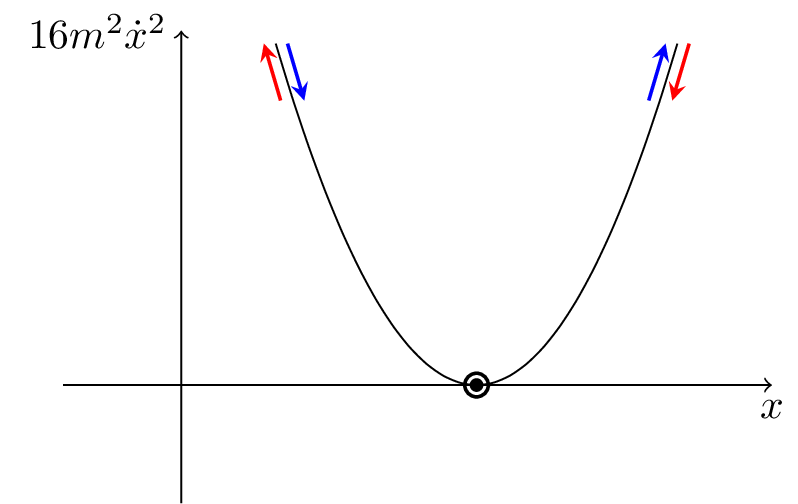} \caption{\label{fig:phasespaceQ4ndelta0}$Q_4<0$ and $\Delta = 0$} \end{subfigure} \begin{subfigure}[t]{0.48\columnwidth} \centering \includegraphics[width=\textwidth]{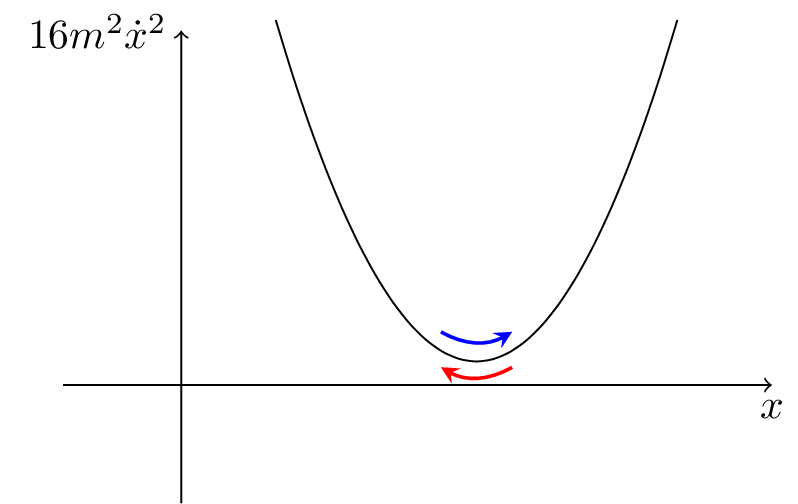} \caption{\label{fig:phasespaceQ4ndeltan}$Q_4<0$ and $\Delta < 0$} \end{subfigure} \begin{subfigure}[t]{0.48\columnwidth} \centering \includegraphics[width=\textwidth]{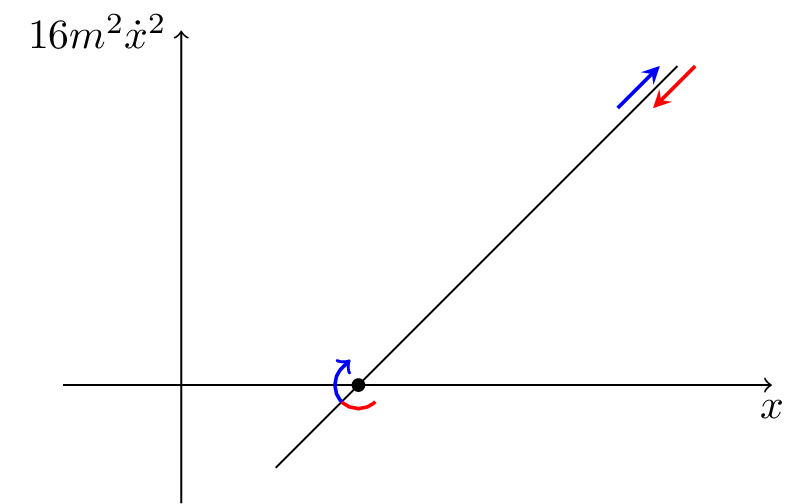} \caption{\label{fig:phasespaceQ40}$Q_4=0$} \end{subfigure} \caption{Trajectories of the vacuum shell in the phase space $(16 m^2 \dot{x}^2, x)$, where $x=\mathcal{R}^2$, for different $Q_4$ and $\Delta$. Trajectories along the blue arrows correspond to the branch with positive velocity, while the ones along red arrows correspond to the branch with negative velocity. The trajectories switch branches at the points with $\dot{x} = 0$ with a curved arrow.\label{fig:phasespace}} \end{figure}

The differential equations in Eq.~\eqref{eq:rdotanalytical} can be analytically integrated, but the branches of integration highly depend on the quantities $Q_0$, $Q_2$, and $Q_4$.
In order to visualize the solutions of Eq.~\eqref{eq:rdotanalytical}, 
we can consider the said equation in the form 
$16 m^2 (\frac{d}{d\tau}(\mathcal{R}^2))^2 = Q_0 + Q_2 \mathcal{R}^2 - Q_4 \mathcal{R}^4$, 
which describes a parabola in the phase space $(16 m^2 \dot{x}^2, x)$, with $x = \mathcal{R}^2$.
The sign of the second derivative of the parabola heavily influences the type of solutions, which is controlled by the sign of the quantity $Q_4$. 
Moreover, the roots of the parabola are given by 
$\mathcal{R}^2_\pm = \frac{Q_2 \pm \sqrt{\Delta}}{2Q_4}$, where $\Delta = Q_2^2 + 4Q_0 Q_4$. 
So, the existence of real roots is controlled by the sign of the quantity $\Delta$. Also, the apex of the parabola resides at $\mathcal{R}^2_a = \frac{Q_2}{2 Q_4}$. Considering these quantities, the trajectories in phase space assume different behaviors (see Fig.~\ref{fig:phasespace}). The types of 
trajectories can be split into oscillating solutions, for $Q_4 > 0$, bouncing and exponential solutions, for $Q_4 < 0$, and a square root solution for $Q_4=0$. There are also static solutions for $\Delta = 0$, which have different stability depending on the sign of $Q_4$.



In the case of $Q_4 > 0$ and $\Delta > 0$, the differential equations in Eq.~\eqref{eq:rdotanalytical} can be integrated to give the solutions 
\begin{equation}
    \mathcal{R} = \sqrt{\frac{Q_2}{2Q_4} + \frac{\sqrt{\Delta}}{2Q_4} \sin \left[\frac{s\sqrt{Q_4}}{4|m|}(\tau+\kappa)\right]}\,,
    \label{eq:oscsolution}
\end{equation}
where $s = \pm 1$ corresponds to the sign of the initial velocity of the shell, 
and $\kappa$ is an integration constant. Note that $\kappa$ must be such that 
$-\frac{\pi}{2}< s\sqrt{Q_4}\kappa/(4m) < \frac{\pi}{2}$ so that an increase in $\tau$ 
leads to the correct behaviour according to the sign of the initial velocity.
The solutions in Eq.~\eqref{eq:oscsolution} have an oscillatory behavior around the radius $\mathcal{R} = 
\sqrt{\frac{Q_2}{2Q_4}}$ and they exist in the bounded region $\mathcal{R}\in [\mathcal{R}_-,\mathcal{R}_+]$, see Fig.~\ref{fig:phasespaceQ4pdeltap}. For $\Delta \leq 0$, there are no dynamic real solutions.

For $Q_4 < 0$, there are three different solutions depending on the sign of $\Delta$ and they can be 
written as
\begin{align}
    &\mathcal{R} = \sqrt{-\frac{Q_2}{2\abs{Q_4}} + k\frac{\sqrt{\Delta}}{2\abs{Q_4}}\cosh\left[\frac{s \sqrt{\abs{Q_4}}}{4|m|k}(\tau + \kappa)\right]},\Delta > 0\,,\label{eq:bottomone}\\
    &\mathcal{R} = \sqrt{-\frac{Q_2}{2\abs{Q_4}} + k \exp\left[\frac{s \sqrt{\abs{Q_4}}}{4|m|k}(\tau+\kappa)\right]}\,\,,\Delta = 0\,,\label{eq:middleone}\\
     &\mathcal{R} = \sqrt{-\frac{Q_2}{2\abs{Q_4}} + \frac{\sqrt{\abs{\Delta}}}{2\abs{Q_4}}\sinh\left[\frac{s \sqrt{\abs{Q_4}}}{4|m|k}(\tau + \kappa)\right]}\,\,,\Delta < 0\,,\label{eq:otherone}
\end{align}
where $k = -1$ if the initial radius satisfies $\mathcal{R}_0 < \sqrt{\frac{-Q_2}{2\abs{Q_4}}}$ and $k=1$ if the initial radius satisfies $\mathcal{R}_0 > \sqrt{\frac{-Q_2}{2\abs{Q_4}}}$. For $\Delta > 0$, the trajectories lie 
in the region $\mathcal{R}\in ]0,\mathcal{R}_-]\cup[\mathcal{R}_+,+\infty[$ (see Fig.~\ref{fig:phasespaceQ4ndeltap}), and they are characterized either by bouncing solutions if $s=-k$ or 
by exponential solutions if $s=k$. For $\Delta =0$, the trajectories lie in the entire domain (see Fig.~\ref{fig:phasespaceQ4pdelta0}). However, if the trajectory starts at $\mathcal{R}_0\neq \sqrt{\frac{-Q_2}{2|Q_4|}}$ with $k=-s$, the shell does not reach $\mathcal{R} = \sqrt{\frac{-Q_2}{2|Q_4|}}$ in a finite time. For $\Delta<0$, we also have trajectories lying in the entire domain and characterized by exponential solutions, see Fig.~\ref{fig:phasespaceQ4ndeltan}.

Finally, for $Q_4 = 0$, the motion of the shell follows the expression
\begin{equation}\label{eq:squareroot}
    \mathcal{R} = \sqrt{\frac{Q_2 (\tau + \kappa)^2}{64m^2} - \frac{Q_0}{Q_2}}\,,
\end{equation}
assuming that $Q_2 \neq 0$. Note that the solution in Eq.~\eqref{eq:squareroot} can also be obtained 
from taking the limit of Eq.~\eqref{eq:bottomone} as $Q_4 \rightarrow 0$. The trajectories in this case 
lie on the interval $\mathcal{R}\in \left[\frac{-Q_0}{Q_2},+\infty\right]$ and they are characterized by a square root solution, see Fig.~\ref{fig:phasespaceQ40}.

\begin{figure}[h]
    \centering
    \includegraphics[width=0.45\textwidth]{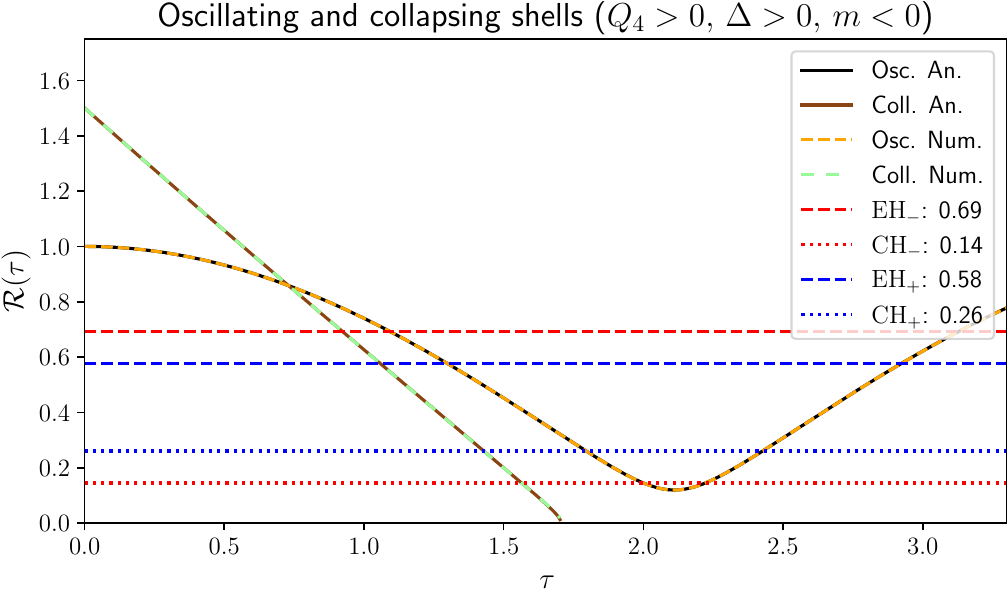}
    \includegraphics[width=0.45\textwidth]{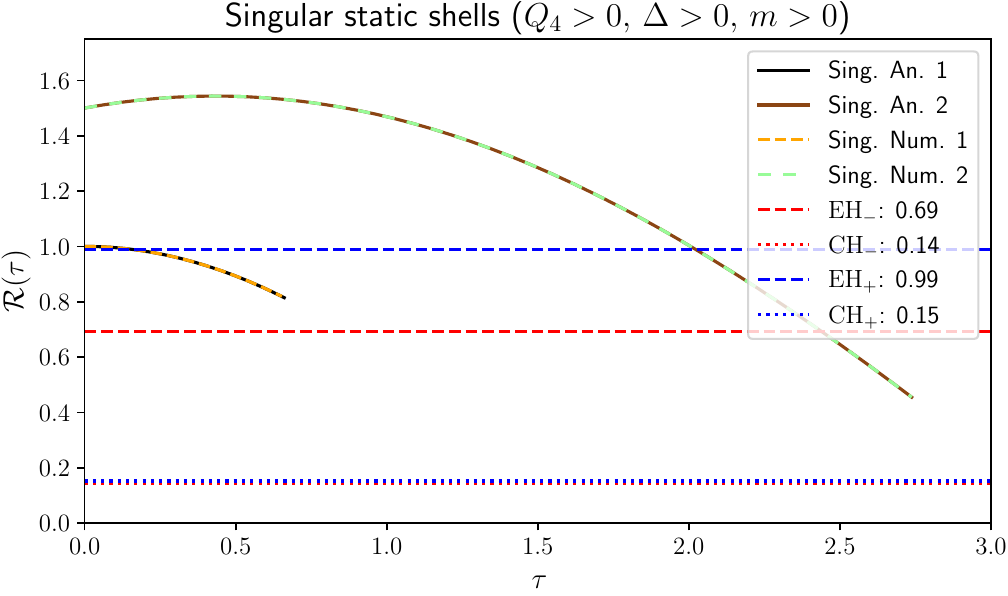}
    \vspace{-3mm}
    \caption{Plots depicting the radius of the thin shell as a function of time $\tau$, when in the regime of Eq.~\eqref{eq:oscsolution}. The top plot shows cases for negative $m$ and with parameters: oscillating shell (Osc.) with 
    $\mathcal{R}_0 = 1.0$ and $\dot{\mathcal{R}}_0 = 0.0$, and collapsing shell (Coll.) with 
    $\mathcal{R}_0 = 1.5$ and $\dot{\mathcal{R}}_0 = -0.9$, with 
    spacetime parameters $M_+ = 0.4$, $j_+ = 0.3$, $M_- = 0.5$, $j_- = 0.2$, 
    together with the corresponding inner and outer event horizons ($\text{EH}_\pm$) and Cauchy horizons ($\text{CH}_\pm$). The bottom plot's parameters are: $M_+ = 1$, $j_+ = 0.3$, $M_- = 0.5$, $j_- = 0.2$, for two 
    singular shells (Sing.), one with $\mathcal{R}_0 = 1.0$ and $\dot{\mathcal{R}}_0 = 0.0$, 
    and another with $\mathcal{R}_0 = 1.5$ and $\dot{\mathcal{R}}_0 = 0.2$. We plot both the analytic solutions (An.), and the numerical evolutions (Num.) in each case.}
    \label{fig:oscsol}
\end{figure}

\begin{figure}[h]
    \centering
    \includegraphics[width=0.45\textwidth]{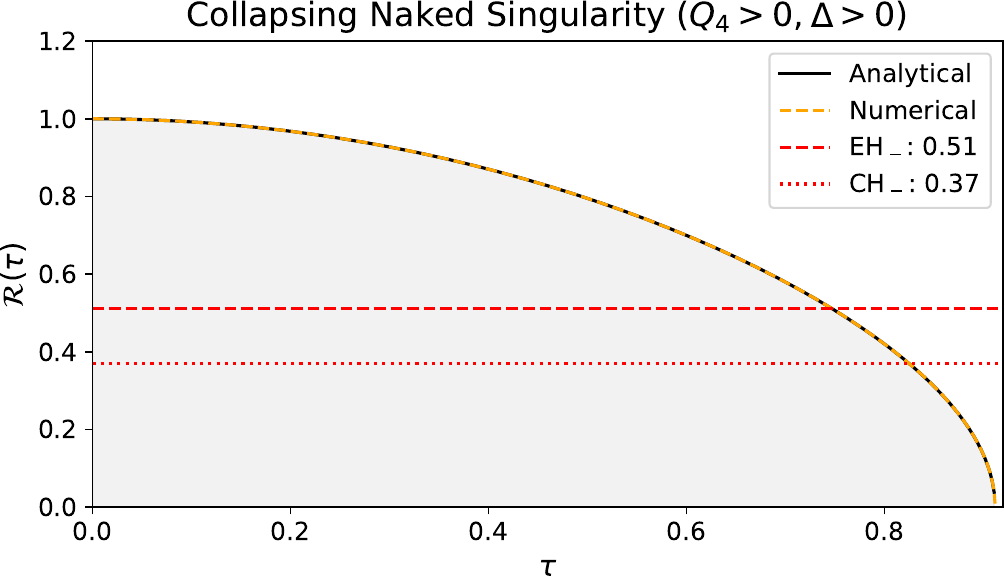}
    \vspace{-3mm}
    \caption{Formation of naked singularity with a negative 
    $m$ vacuum thin shell, with spacetime parameters $M_+ = 0.4$, $j_+=0.5$, $M_-=0.4$ 
    and $j_-=0.38$. We shadow the region under the curve to make the separation of the two spacetimes clearer.}
    \label{fig:nakedsing}
\end{figure}

\begin{figure*} 
\centering 
\begin{subfigure}[t]{0.49\textwidth} 
\centering \includegraphics[width=\textwidth]{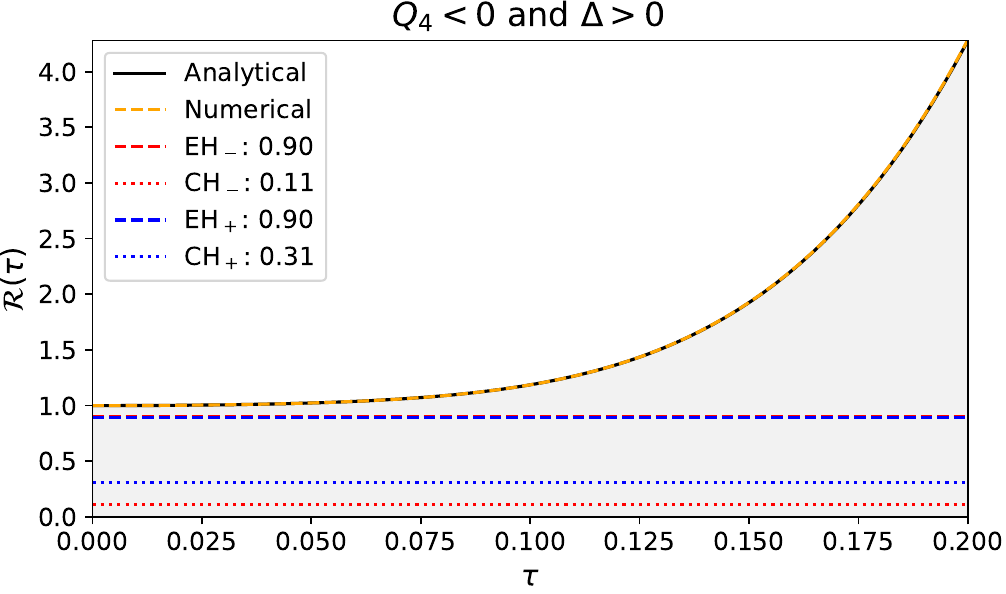} \end{subfigure} \begin{subfigure}[t]{0.49\textwidth} \centering \includegraphics[width=\textwidth]{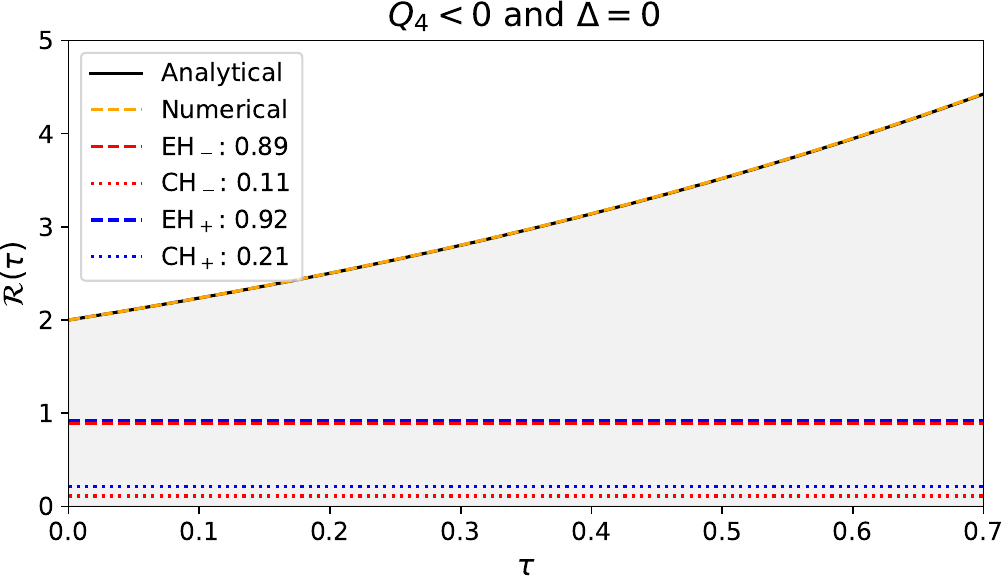} \end{subfigure}\\ \begin{subfigure}[t]{0.49\textwidth} \centering \includegraphics[width=\textwidth]{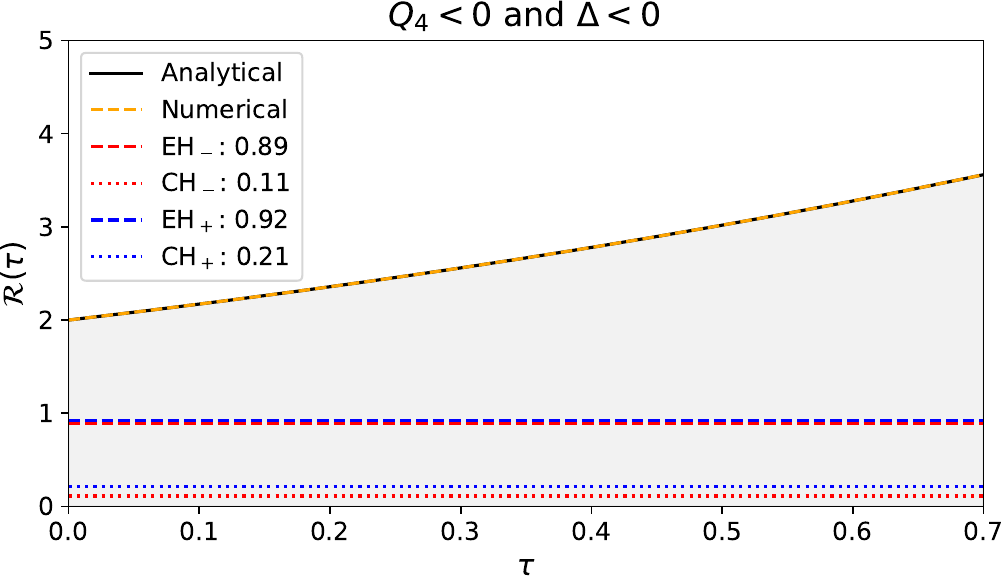}\end{subfigure} \begin{subfigure}[t]{0.49\textwidth} \centering \includegraphics[width=\textwidth]{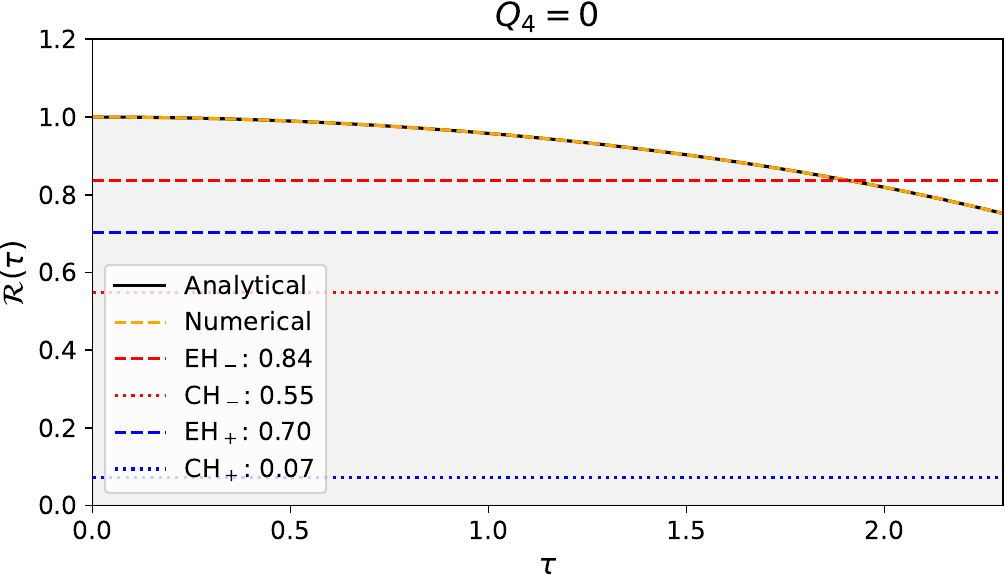} \end{subfigure}\caption{Examples of the behaviors described by Eqs.~\eqref{eq:bottomone} (top left), \eqref{eq:middleone} (top right), \eqref{eq:otherone} (bottom right), and \eqref{eq:squareroot} (bottom left). We show both the analytic predictions due to the equations ("Analytical"), and the numerical integration of $\dot{m} = 0$ ("Numerical"). In order, the sets of parameters that describe each graph are: (top left) $M_+ = 0.9$, $M_- = 0.83$, $j_+ = 0.56$, $j_- = 0.2$, $s = -1$, $\mathcal{R}_0 = 1$, $\dot{\mathcal{R}}_0 = 0$; (top right) $M_+ = 0.9$, $M_- = 0.8$, $j_+ = 0.39$, $j_- = 0.2$, $s = 1$, $\mathcal{R}_0 = 2$, $\dot{\mathcal{R}}_0 = 2.2261$; (bottom left) $M_+ = 0.9$, $M_- = 0.8$, $j_+ = 0.39$, $j_- = 0.2$, $s = 1$, $\mathcal{R}_0 = 2$, $\dot{\mathcal{R}}_0 = 1.64$; (bottom right) $M_+ = 0.5$, $M_- = 1.0$, $j_+ = 0.1$, $j_- = 0.917745$, $s = 1$, $\mathcal{R}_0 = 1$, $\dot{\mathcal{R}}_0 = 0$.\label{fig:restofexamples} } \end{figure*}

For specific cases, we have compared the analytic solutions with the numerical 
evolution of Eq.~\eqref{eq:secondEquatioOfMotion}. For all cases we have 
compared, the analytic solution is consistent with the numerical evolution. 
In particular, the cases for $Q_4>0$ are displayed in Fig.~\ref{fig:oscsol}, and
the cases for $Q_4\leq0$ are displayed in Fig.~\ref{fig:restofexamples}. 

The case presented in Fig.~\ref{fig:nakedsing} depicts the collapse of 
a vacuum thin shell with negative $m$ into a naked singularity. Since the inner 
spacetime at beginning has a horizon, this means that there is a formation of a naked 
singularity. The existence of naked singularities is not new in EGB, having been reported in Refs.~\cite{garraffo_lovelock_2008, Dotti:2007az,nozawa_effects_2006} previously, but this seems to be one of the first examples of the dynamical creation of a naked singularity in said theory.

Note that for all the analytic solutions above, one must check when the
continuation occurs, i.e. when the analytic solution ceases to obey the 
differential equation in Eq.~\eqref{eq:secondEquatioOfMotion} 
precisely at the points obeying $A_\pm + \dot{\mathcal{R}}^2 = 0$, which generally 
occurs inside the largest horizon.
The analytic solutions must then be truncated at these points. While a full 
analysis in terms of spacetime parameters is quite involved, the examples 
we have checked are compatible with this cutoff occurring for positive $m$ but not on 
negative $m$. Furthermore, the cutoff happens independently of the values of 
$Q_4$.

\subsubsection{Static solutions}

In addition to the dynamical solutions, there are also static solutions for the vacuum shell, which are illustrated in Fig.~\ref{fig:phasespace}. These solutions occur when $\dot{\mathcal{R}} = 0$ and $\ddot{\mathcal{R}}=0$, which in the phase space are represented by the apex of the parabola, $\mathcal{R}=\sqrt{\frac{Q_2}{2Q_4}}$, for $\Delta = 0$. It can be shown that indeed these two conditions together are consistent with $p_\mathcal{R} = 0$, i.e., it satisfies 
\begin{equation}\label{eq:pR0}
    \frac{\mathcal{R} A_+^\prime + 2 A_+}{\sqrt{A_+}} = \frac{\mathcal{R}A_-^\prime + 2 A_-}{\sqrt{A_- }}\,.
\end{equation}
The condition $\Delta = 0$ can be cast into a quadratic polynomial equation for $m^2$, whose roots are $m_{1,2}^2$, written in terms of the spacetime parameters. Plugging these roots into $\mathcal{R}=\sqrt{\frac{Q_2}{2Q_4}}$, we obtain the two possible values of the radius, $\mathcal{R}_{1,2}$ that solve Eq.~\eqref{eq:pR0}. The radii $\mathcal{R}_{1,2}$ and the parameters $m_{1,2}$ of the static solutions in terms of the spacetime parameters are 
\begin{align}
    &\mathcal{R}_{1,2} = \sqrt{\frac{M_-(M_+^2 - j_+^2)\hskip-1mm-\hskip-1mm M_+(M_-^2 - j_-^2) \hskip-1mm \mp\hskip-1mm (M_+\hskip-1mm -\hskip-1mm M_-)\xi}{2(M_+^2-j_+^2 - M_-^2 + j_-^2)}}\,,\\
    &m^2_{1,2} = \frac{1}{4}\left[j_-^2+j_+^2-2M_-M_+\pm \xi \right],
    \label{eq:stationary}
\end{align}
 with 
 \begin{align}
     \xi = \sqrt{( M_-^2 -j_-^2)( M_+^2-j_+^2)}\,.
 \end{align}
Note that for the radius $\mathcal{R}_1$, the mass parameter of the shell is $m_1$, while for the radius $\mathcal{R}_2$, the mass parameter is $m_2$.

Now, regarding the stability of the static solutions $\mathcal{R}_{1,2}$, it is quite involved to understand the condition Eq.~\eqref{eq:stability} in terms of the phase space. Instead, we can consider perturbations in Eq.~\eqref{eq:rdotanalytical}, or more precisely in its derivative 
\begin{align}\label{eq:rdotdot2}
    \ddot{\mathcal{R}} = -\frac{Q_0 + Q_4 \mathcal{R}^4}{(8m)^2 \mathcal{R}^3}\,\,.
\end{align}
In order to perform a perturbation in $\mathcal{R}$, we must understand how $m$ can change. Indeed, changing the initial conditions slightly may change $m$, although $m$ is a constant of motion. Under a change of $\mathcal{R}$, we get
\begin{align}
    &\delta m = - (8\pi p_\mathcal{R} \mathcal{R})|_{0} \delta \mathcal{R}\notag\\
    &- \eval{\left(\frac{\dot{\mathcal{R}}m}{\sqrt{A_-+\dot{\mathcal{R}}^2}\sqrt{A_++\dot{\mathcal{R}}^2}}\right)}_0 \delta \dot{\mathcal{R}}\,\,,
\end{align}
where the subscript $0$ means it is evaluated at the static solution, and we are keeping everything at 
first-order. However, the static solutions imply $p_\mathcal{R}=0$ and $\dot{\mathcal{R}}=0$, so $\delta m = 0$ at first order. This simplifies the analysis heavily, since we can perform the variation of $\mathcal{R}$ in Eq.~\eqref{eq:rdotdot2} while leaving the phase space parameters fixed. Indeed, we have
\begin{align}
    \delta \ddot{\mathcal{R}} = - \frac{4 Q_4}{(8m)^2}\delta \mathcal{R}\,\,,
\end{align}
hence the static solutions with $Q_4> 0$ are stable, while the static solutions with $Q_4<0$ are unstable. 

The stability can be visualized in the phase space diagram as follows. When $Q_4 > 0$, a small deviation from the static solutions leads to an increase in $\Delta$, and so the diagram in Fig.~\ref{fig:phasespaceQ4pdelta0} turns into Fig.~\ref{fig:phasespaceQ4pdeltap}, which has an oscillatory behaviour, marking stability. When $Q_4 < 0$, a small deviation from the static solutions 
may increase or decrease $\Delta$, in any case, the shell goes away from the equilibrium radius exponentially at late times. In the diagram, this means the shell at the apex of Fig.~\ref{fig:phasespaceQ4ndelta0} turns into either Figs.~\ref{fig:phasespaceQ4ndeltap} or~\ref{fig:phasespaceQ4ndeltan}.

\begin{figure}
    \centering
    \includegraphics[width=1\linewidth]{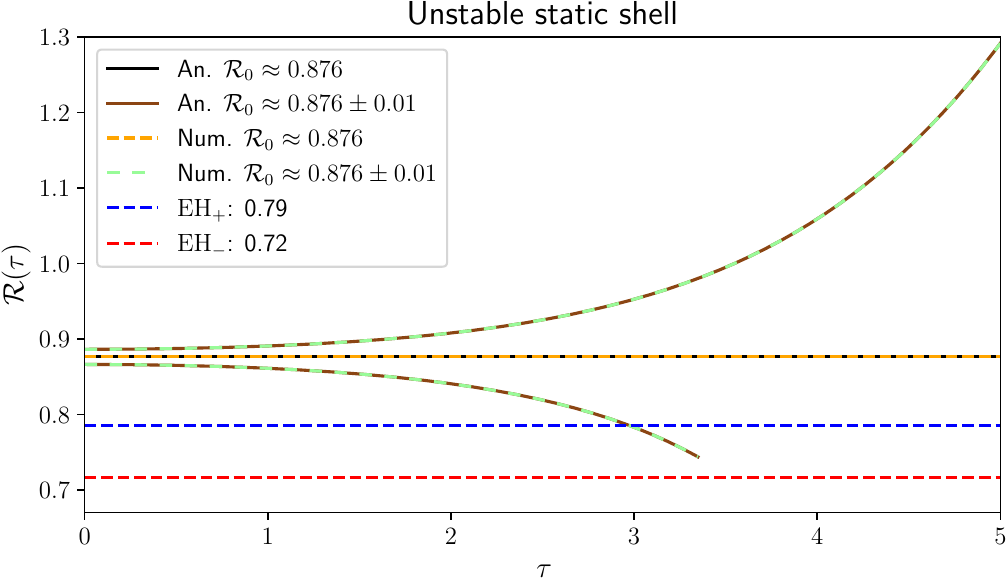}
    \caption{Unstable static shell $\mathcal{R}_0\approx 0.876$ for spacetime parameters 
    $M_+ = 1.01$, $j_+=0.985$, $M_-=0.65$ and $j_-=0.53$. To illustrate 
    instability, the evolution of shells with $\mathcal{R}_0\approx 0.876\pm 0.01$ are also displayed.}
    \label{fig:unstable}
\end{figure}

\begin{figure}
    \centering
    \includegraphics[width=1\linewidth]{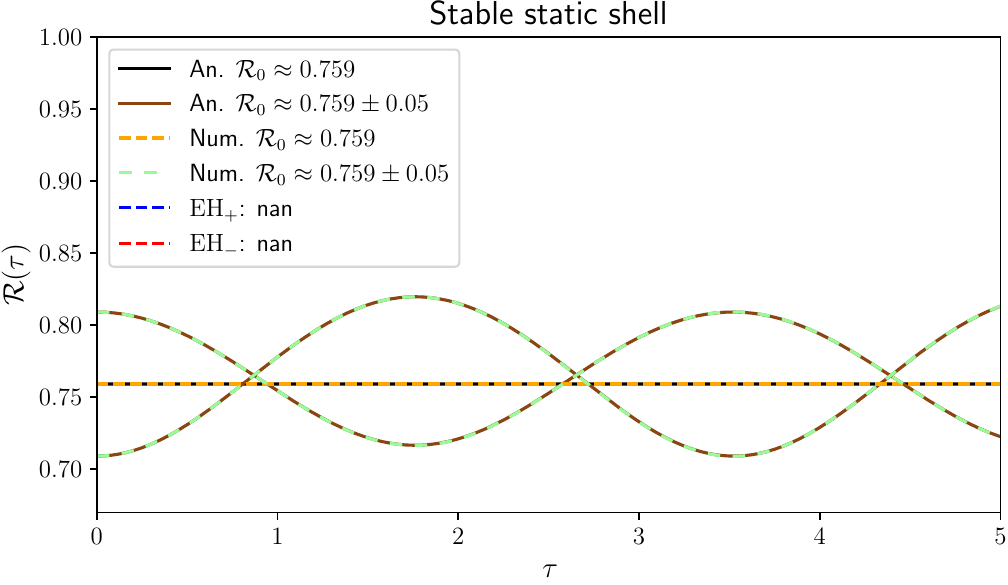}
    \caption{Stable static shell $\mathcal{R}_0\approx 0.759$ for spacetime parameters 
    $M_+ = 1.01$, $j_+=1.05$, $M_-=0.36$ and $j_-=1.635$. To illustrate 
    instability, the evolution of shells with $\mathcal{R}_0\approx 0.759\pm 0.05$ are also displayed.}
    \label{fig:stable}
\end{figure}

In Figs.~\ref{fig:unstable} and~\ref{fig:stable}, we show particular spacetime parameters 
where these static thin shells occur. While a full scan of parameters was not performed, 
the quick analysis of the function $m$ tells us that the unstable static thin shells 
occur when the event horizons from the inner and outer spacetimes approach each other 
and when they get close to extremality. The stable static thin shells occur when both 
inner and outer spacetimes are overextremal. 

\subsection{The case of $b_\pm \neq 0$}
When $b_\pm \neq 0$, no analytical solution was found. The difference in terms of integrating Eq.~\eqref{eq:rdotanalytical} is the fact that now terms proportional to $\mathcal{R}$ and to $\mathcal{R}^3$ show up. An interesting scenario in this case is setting the interior spacetime to be the vacuum associated with this theory, which is not $\text{AdS}_5$. Evolving numerically Eq.~\eqref{eq:rdotdot}, we report the possibility of collapsing naked singularities again. In general, the oscillating solutions seem not to exist in this regime, which agrees with the added nonlinearities that come from $b_\pm \neq 0$. 

\section{Conclusions}
\label{sec:conclusions}

In this work, we have studied thin shells gluing two spacetimes described by the 
metric in~\cite{Anabalon:2024abz} in the Chern-Simons point of Einstein Gauss Bonnet 
gravity in five dimensions. In order to glue both spacetimes, we proposed 
an anisotropic stress tensor for the shell and we observed that the only permissible 
shells have a non-zero pressure in the $\rho$ direction, with the other components 
vanishing, or they can also have a vanishing stress tensor. We then analyzed and 
obtained analytical solutions for vacuum thin shells. A property of these shells is 
their parameter $m$, which plays a role of a mass parameter but it is not linked 
to the surface stress tensor, as the latter vanishes. This unusual behaviour 
raises various questions about the physical origin of $m$. This can be a 
feature of the Davis junction conditions, as the parameter 
$m$ may be related to the higher order curvature terms. Indeed, 
there is an intrinsic discontinuity of the extrinsic curvature of the shell, sourced 
by the curvature itself. However, further inspection 
is needed to ascertain why these shells are permitted, given that 
in GR they are connected to the surface matter density. Another topic related to the 
junction conditions is that, from lessons of GR, we expected two conditions 
when gluing two rotating spacetimes: one describing the difference in the mass 
of both spacetimes and another describing the difference in the angular momentum. 
However, here we just have the condition for $m$ and so there is additional freedom 
in the parameters of the rotating spacetimes. It is unclear if this is a feature 
of CS EGB and it might be an indication that one should be very careful about the 
relevant physics that one can extract from these rotating solutions.

The analytic expressions for the motion of the vacuum thin shell 
were obtained by reducing the order of the equation of motion through 
the constant of motion $m$. The expressions then depend on combinations 
of the spacetime parameters and $m$. We described the different types of 
motion depending on the values of these combinations, the quantities 
$Q_0$, $Q_2$ and $Q_4$. It would be tentative 
to describe analytically the regions 
in terms of the initial value parameters where these types of motion occur, 
however the analysis becomes quite involved due to the number of 
variables and the polynomial order of the combinations. While some statements 
can be made by looking at $m$ in function of the spacetime parameters and the initial 
radius, with zero velocity, it is unclear if a full analytic description 
can be made. Such a description would be important for example to understand 
the precise range of parameters where the static thin shells exist.
We recognize that the 
description of the different regimes in terms of the combinations 
of spacetime parameters becomes illusive but this may be the price to pay 
for getting analytic expressions that solve the non-linear partial differential 
equation. 

In the examples we have shown, there are points where the differential equation 
becomes ill-defined. The analytic solutions found are able to perform a continuation 
beyond such points, however $m$ ceases to be constant. At these points, some interesting features arise. If one describes the shell by $t=\mathcal{T}(\tau)$ and 
$r=\mathcal{R}(\tau)$, the junction conditions lead to $\dot{\mathcal{T}} = 0$ while 
$\dot{\mathcal{R}}(\tau)$ is some non-zero value. In terms of outer spacetime coordinates, this leads to $t=\mathrm{const}$. Moreover, while the normal vector 
is still well defined, the extrinsic curvature as seen from 
the outer spacetime diverges. A singular extrinsic curvature may indicate that 
the metric ceases to be continuous. Beyond this statement, it is not clear how the 
evolution carries on or what happens to the vacuum shell.

The possibility of producing naked singularities is not new in EGB, and here we were 
able to show that they can be dynamically produced in this theory. This was possible 
due to the implementation of the thin shell formalism, where the inner and outer 
spacetimes are described by the known rotating solution in EGB \eqref{eq:Anabalon}, 
valid only for the Chern-Simons point. One of the drawbacks of Eq.~\eqref{eq:Anabalon} 
is that it cannot be compared to the Boulware-Deser solution, in the limit of slow 
rotation, or to the Kerr solution, in the limit of small $\alpha$. On the other hand, 
this solution of the Chern-Simons point allows for fundamentally different physics from 
usual GR, and even the usual EGB. Whether it is a relevant issue to pursue further, we 
shall relay this discussion for future work.

\section*{Acknowledgements}
The authors thank Jorge V. Rocha, Antonia Frassino, and José Natário for helpful discussions leading to this work. 
The authors extend the acknowledgments to José M. M. Senovilla and Raül Vera for the useful explanations of the problems in deriving the Einstein-Gauss-Bonnet junction conditions. JDA thanks Fundação para a Ciência e Tecnologia (FCT), Portugal, for the financial support through the Grant Project  2024.04456.CERN. TVF thanks the Fundação para Ciência e Tecnologia (FCT), Portugal, 
for the financial support to the Center for Astrophysics and Gravitation (CENTRA/IST/ULisboa) through the grant No. UID/PRR/00099/2025 and 
grant N0. UID/00099/2025.

\bibliography{references}

\newpage

\end{document}